\documentclass[pss,fleqn,embeddedheads]{w-art}
\usepackage{times}
\usepackage{w-thm}
\usepackage[]{graphicx}
\begin{document}
\DOIsuffix{theDOIsuffix}
\Volume{(c) 1}
\Issue{1}
\Copyrightissue{01}
\Month{01}
\Year{2004}
\pagespan{1}{}
\Receiveddate{\sf 1 September 2003} \Reviseddate{\sf 16 September 2003} \Accepteddate{\sf
16 September 2003} \Dateposted{\sf 28 November 2003}



\title[Manuscript preparation guidelines]{Electron localization in an
external electric field}


\author[O. Bleibaum]{O. Bleibaum\footnote{Corresponding
     author: e-mail: {\sf olaf.bleibaum@physik.uni-magdeburg.de},
Phone: +49\,391\,67\,12474, Fax:
     +49\,391\,67\,11217}\inst{1}}
\address[\inst{1}]{Institut f\"ur Theoretische Physik, Otto-von-Guericke
Universit\"at Magdeburg, 39016 Magdeburg, PF 4120, Germany}
\author[D. Belitz]{D. Belitz\inst{2}}
\address[\inst{2}]{Department of Physics and Materials Science Institute,
University of Oregon, Eugene OR 97403, USA} 
\begin{abstract}
The impact of a weak electric field on the weak-localization corrections is
studied within the framework of a nonlinear $\sigma$-model. Two scaling regimes
are obtained. In one, the scaling is dominated by temperature; in the other, by
the electric field. An explicit expression is derived for the crossover
temperature between the two regimes.
\end{abstract}
\maketitle                   




\renewcommand{\leftmark}
{O. Bleibaum and D. Belitz: Localization in an electric field}
\section{Introduction}
The physics of weakly disordered systems has been the subject of considerable
interest over the past years. If interactions effects can be ignored, it is
known that quantum interference effects lead to the localization of all
electronic states in two-dimensions, even for arbitrarily weak disorder
\cite{LeeRama}. That is, the ohmic conductivity of such systems vanishes. It is
well understood how the localization is affected by external fields that
destroy the phase coherence underlying the quantum interference effects, e.g.,
magnetic fields \cite{LeeRama}. However, there still are open questions about
the impact of an electric field, which does not break time reversal invariance,
and thus has no direct impact on the interference effects. This is an important
question, since transport experiments invariably involve electric fields. In
the theoretical literature, one can find arguments that a weak electric field
has no impact on the localization corrections at all \cite{Altshuler}, to
arguments that predict a strong impact on the localization corrections already
at very small fields \cite{Tsuzuki}, to arguments that predict immediate
delocalization \cite{Kirkpatrick}, to arguments that predict delocalization
beyond a critical field strength \cite{Bryksin}. Here we revisit this question.
In agreement with Ref.\ \cite{Kirkpatrick}, we find that an arbitrarily weak
field destroys the localization. However, for weak disorder this effect is not
observable at realistic temperatures.

\section{The model}
We consider a Hamilton operator
\begin{equation}\label{I1}
H = \frac{{\hat{\text{\bf\textit p}}}^2}{2m} + \text{\bf\textit
F}\cdot\text{\bf\textit x} + V(\text{\bf\textit x}).
\end{equation}
Here ${\hat{\text{\bf\textit p}}}$ is the momentum operator, $m$ is the
electron mass, $\text{\bf\textit x}= (x,y)$ is the position in real space,
$\text{\bf\textit F} = (F,0)$ is the electric field, and $V(\text{\bf\textit
x})$ is a random potential. For the latter we assume a Gaussian distribution
with zero mean and a second moment given by
\begin{equation}\label{I2}
\langle V(\text{\bf\textit x})V(\text{\bf\textit y})\rangle =
\frac{\hbar}{2\pi\nu\tau}\ \delta(\text{\bf\textit x} - \text{\bf\textit y}),
\end{equation}
where $\nu$ is the density of states and $\tau$ is the relaxation time in the
self-consistent Born approximation.

To investigate the impact of the field on the electron localization, we focus
on the density relaxation, as described by the integral equation
\begin{equation}\label{I2a}
n(\text{\bf\textit x}, E|\Omega)= \int d\text{\bf\textit y}\ P(\text{\bf\textit
x}, \text{\bf\textit y}|E,\Omega)\,
   n_0(\text{\bf\textit y}, E).
\end{equation}
Here $n_0(\text{\bf\textit x},E)$ is the initial distribution of the number of
particles with energy $E$ at position $\text{\bf\textit x}$, the propagator $P$
describes the evolution of this distribution, and $n(\text{\bf\textit
x},E|\Omega)$ is the Laplace transform of the time dependent particle number
density $n(\text{\bf\textit x},E|t)$.

To calculate the density propagator $P$ we generalize the nonlinear
$\sigma$-model of Ref.\ \cite{Wegner} to allow for the presence of a weak
electric field. The action takes the form
\begin{equation}\label{I6}
S = -\frac{\pi\nu\hbar}{4} \int d\text{\bf\textit x} \hskip -5pt
           \sum_{\alpha=-n_r+1}^{n_r} \Omega\,
                   Q_{\alpha\alpha}(\text{\bf\textit x})\, \Lambda_{\alpha}
    - \frac{\pi\nu\hbar}{8}\int d\text{\bf\textit x} \hskip -10pt \sum_{\alpha,\alpha'=-n_r+1}^{n_r}
       \hskip -5pt Q_{\alpha\alpha'}(\text{\bf\textit x})\, [{\bf \nabla}\cdot D(\mu_x){\bf\nabla}]\,
           Q_{\alpha'\alpha}(\text{\bf\textit x}).
\end{equation}
Here $\alpha$ numbers $2n_r$ replicas, $\Lambda_{\alpha}=+1$ if $\alpha>0$, and
$\Lambda_{\alpha}=-1$ if $\alpha\leq 0$. The matrix field $Q$ is defined on a
$O(n_r,n_r)/O(n_r)\times O(n_r)$ Grassmann manifold and has the properties
$Q^2=1$ and $\mbox{tr}\,Q=0$. The integration over real space is understood to
be restricted to the classically accessible region, defined by the requirement
$\mu_x \equiv E-\text{\bf\textit F}\cdot \text{\bf\textit x} > 0$. The
generalized diffusion coefficient is given by
\begin{equation}\label{I7}
D(\mu_x) = \tau\mu_x/m.
\end{equation}
In deriving Eq.\ (\ref{I6}) we have restricted the consideration to small
fields, which satisfy the relationship $Fl/\mu_x\ll 1$, where $l$ is the mean
free path. It can be shown that the two-point propagator of this field theory
determines $P$.

To investigate the action (\ref{I6}) we parameterize the $2n_r\times 2n_r$
matrix $Q$ in terms of real $n_r\times n_r$ matrices $q$, according to
\begin{equation}\label{I15}
Q=\left(
\begin{array}{cc}
\sqrt{1+qq^T}&q\\
-q^T&-\sqrt{1+q^Tq}
\end{array}
\right),
\end{equation}
and expand the action (\ref{I6}) in powers of $q$. For a one-loop calculation
it suffices to keep terms up to $O(q^4)$.


\section{The Gaussian fluctuations}
We first consider the Gaussian approximation. The action (\ref{I6}) then yields
a generalized diffusion equation for $P$,
\begin{equation}\label{I8}
\left(\Omega-[{\bf \nabla}\cdot D(\mu_x){\bf
\nabla}]\right)\,P(\text{\bf\textit x}, \text{\bf\textit y}|E,\Omega)=
\delta(\text{\bf\textit x} - \text{\bf\textit y}).
\end{equation}
This differential equation must be supplemented by boundary conditions. We
require that the propagator vanishes at infinity in the classically accessible
region, and that the current in the direction of the field vanishes at the
classical turning point. The structure of the equation is a consequence of the
symmetries of the action. It reflects the fact that the configuration averaged
Green functions are symmetric with respect to exchange of $\text{\bf\textit x}$
and $\text{\bf\textit y}$, and invariant against generalized real-space
translations $\text{\bf\textit x}\to\text{\bf\textit x}+\text{\bf\textit a}$,
$E\to E+\text{\bf\textit F}\cdot\text{\bf\textit a}$.

The general solution of Eq.(\ref{I8}) can be expressed in terms of
hypergeometric functions. For our purposes, it is more illuminating to consider
a special initial condition, namely, a homogeneous density $n$ of charge
carriers, all with energy $\mu$. An electric field is then suddenly switched on
at time $t=0$, so
\begin{equation}\label{I5}
n_0(\text{\bf\textit x},E) = n\,\delta(E-(\mu+\text{\bf\textit F}\cdot
\text{\bf\textit x})).
\end{equation}
For such an initial charge carrier density the solution of Eq.(\ref{I8}) gives
the probability for finding the charge carriers at time $t$ with energy $\mu'$
if they had the energy $\mu$ at time $t=0$,
\begin{equation}\label{I9}
{\cal P}(\mu',\mu|t) = \frac{\theta(\mu)\theta(\mu')}{D'F^2t}\
\exp(-\frac{\mu+\mu'}{D'F^2t})\ I_0(\frac{2}{D'F^2t}\sqrt{\mu\mu'}),
\end{equation}
where $I_0$ is the modified Bessel function, and $D'=\tau/m$ is the Drude
mobility. The first moment of this distribution, that is, the mean energy,
increases with time according to
\begin{equation}\label{I10}
\epsilon_{\mu}(t) \equiv \int_0^{\infty}d\mu'\ \mu'\,{\cal P}(\mu',\mu|t) = \mu
   + D'F^2t.
\end{equation}
This heating is accompanied by an ohmic current,
\begin{equation}\label{I11}
\text{\bf\textit j} = -D'n\text{\bf\textit F}.
\end{equation}
The generalized diffusion equation (\ref{I8}) thus describes heating of the
charge carriers due to the work done on the system by the electric field. Eq.\
(\ref{I11}) also shows that $D'$ is indeed the mobility.

Fluctuations about the mean energy, calculated from the equation
\begin{equation}\label{I11a}
\sigma_{\mu}^2(t) = \int_0^{\infty}d\mu'\
         (\mu'-\epsilon_{\mu}(t))^2\,{\cal P}(\mu',\mu|t),
\end{equation}
increase with time according to
\begin{equation}\label{I12}
\sigma_{\mu}^2(t) = (D'F^2t)^2+2D(\mu)F^2t.
\end{equation}
Therefore, deviations from the mean energy are only negligible for small times,
$t\ll t^{*}$, where
\begin{equation}\label{I13}
t^{*} = 2\mu^2/D(\mu)F^2.
\end{equation}
For $t\gg t^{*}$ the fluctuations about the mean energy are as large as the
mean energy itself, so that the mean energy does no longer describe the state
of the system adequately.

A more detailed analysis of the solution shows that $t^{*}$ also sets the time
scale for a change of the structure of a particle packet. At $t=0$ the packet
is a delta pulse in energy space, and for $t\ll t^*$ its spread is Gaussian. In
this limit, the width of the particle packet increases with time in the same
way as in the absence of the field. Therefore, the diffusion volume is not
affected by the field for $t\ll t^{*}$. However, at $t\approx t^*$ the particle
packet undergoes a restructuring. For $t\gg t^*$, the mean square deviation in
the direction of the electric field increases with time according to
\begin{equation}\label{I14}
\langle (x-\langle x\rangle)^2\rangle\approx (D'Ft)^2.
\end{equation}
Here $\langle\dots\rangle$ denotes an average with respect to the distribution
${\cal P}$. For $t\gg t^*$, the diffusion volume thus increases much faster
with time than in the absence of the field.


\section{Scaling, and weak-localization corrections}

We now perform a scaling analysis of the action (\ref{I6}), using the technique
of Ref.\ \cite{Belitz}. To this end we extend the dimensionality of the system
from 2 to $d=2+\epsilon$, and first consider the Gaussian fixed point that
describes a diffusive phase \cite{Belitz}. At this fixed point, the Gaussian
action is invariant against changes of scale of the form $q\to q'=b^dq$,
$\Omega\to \Omega'=b^2\Omega$ and $x'=x/b$, if the electric field is scaled
according to $F\to F'=Fb$. Accordingly, $F$ is a relevant operator with respect
to the diffusive fixed point, with a scale dimension of 1. If we assign a
coupling constant $u$ to the terms quartic in $q$, we find that $u$ decreases
according to $u\to u'= b^{-(d-2)}u$, so $u$ is an irrelevant operator with
respect to the diffusive fix point for $d>2$, and so are all terms of higher
order in $q$. Since the scale dimension of the diffusion coefficient $D$ at the
diffusive fix point is zero, we have the scaling equation
\begin{equation}\label{I15a}
D(\Omega,F,u)=D(\Omega b^2,Fb,ub^{-(d-2)}).
\end{equation}
Putting $\Omega=0$, and $b=1/F$, we find
\begin{equation}\label{I16}
D(0,F,u)\propto {\mbox{const}}+F^{d-2}.
\end{equation}
Equation (\ref{I16}) suggests that in two-dimensions the weak-localization
correction to the diffusion coefficient are logarithmic with respect to $F$. We
have verified this by an explicit calculation of the one-loop corrections. The
corrections obtained in this way are the same as those obtained in Ref.\
\cite{Tsuzuki}.

For $\Omega\neq 0$ there are two scaling regimes. For small $\Omega$ the
scaling of the diffusion coefficient is governed by the field, and for large
$\Omega$, by the frequency. The scaling analysis shows that the crossover
between these scaling regimes occurs at $\Omega\approx \Omega^{*}$, where
$\Omega^{*}\sim F^2$. As one would expect from these arguments, the explicit
calculation of the one-loop corrections yields $\Omega^{*}=\kappa/t^{*}$, where
$\kappa = O(1)$.

The critical fixed point, which describes the Anderson localization transition
in the absence of an electric field, is more complicated. It cannot be found by
power counting alone, but rather requires an explicit calculation within the
framework of a loop expansion \cite{LeeRama,Belitz}. In particular, the scale
dimension of $F$ is determined by the loop expansion. However, in
$d=2+\epsilon$ the scale dimensions at the diffusive and the critical fixed
points, respectively, differ only by terms of $O(\epsilon)$. The leading
contribution to the scale dimension $[F]$ of $F$ is therefore still given by
power counting, and we have
\begin{equation}
[F] = 1 + O(\epsilon).
\label{last}
\end{equation}
The electric field is thus a relevant operator with respect to the Anderson
localization fixed point. Strictly speaking, this discussion shows only that
the localization fixed point is unstable in the presence of an electric field,
and it does not tell what happens instead. However, explicit perturbative
calculations to one-loop order suggest that there is a metallic phase in $d=2$
for $F\neq 0$.

\section{Conclusions}
The arguments presented above show that in a weak electric field the scaling of
the dc-conductivity at asymptotically low temperatures is governed by the
electric field. However, any experiment effectively measures the conductivity
at a frequency or temperature given by the inverse phase relaxation time,
$\tau_{\phi}^{-1}$. Therefore, the impact of the field on the weak-localization
corrections can only be observed if $\Omega^{*}\tau_{\phi}\gg 1$. Quantitative
estimates show $\Omega^{*} < 10^4$Hz, while a typical phase relaxation rate at
dilution refrigerator temperatures is on the order of $10^{11}$Hz. The field
scaling is therefore not observable in current experiments.

These considerations solve the following paradox. While it is true that an
arbitrarily small electric field destroys localization, as was found by
Kirkpatrick \cite{Kirkpatrick}, this effect manifests itself only at
unobservably low temperatures. This explains why the experimental results are
consistent with the zero-field theory, even though electric fields are present
in the transport experiments.

We finally note that in our model inelastic collisions are not taken into
account. Therefore, our results apply only to samples that are shorter than the
energy relaxation length. The consideration of inelastic collisions is
important in order to establish a nonequilibrium steady state characterized by
an effective electron temperature. In our present treatment such a steady state
is absent; the charge carriers are continuously heated up. We therefore plan
further investigations that will focus on energy relaxation processes.
\begin{acknowledgement}
We would like to thank V.~V. Bryksin and T.~R. Kirkpatrick for helpful
discussions. This work was supported by the DFG under grant No. Bl456/3-1, and
by the NSF under grant No. DMR-01-32555.
\end{acknowledgement}

\end{document}